\documentclass[prl,twocolumn,showpacs]{revtex4}
\usepackage{graphicx}

\begin{document}
\title{Spin-wave scattering at low temperatures in manganite films }
\author{X. J. Chen$^{1}$, H.-U. Habermeier$^{2}$, C. L. Zhang$^{1}$, H. Zhang$^{2}$, and C.
C. Almasan$^{1}$ }
\affiliation{$^{1}$Department of Physics, Kent State University, Kent, Ohio 44242\\
$^{2}$Max-Planck-Institut f\"{u}r Festk\"{o}rperforschung, D-70569 Stuttgart, Germany }
\date{\today}

\begin{abstract}
The temperature $T$ and magnetic field $H$ dependence of the resistivity $\rho$ has been
measured for La$_{0.8-y}$Sr$_{0.2}$MnO$_{3}$ (y=0 and 0.128) films grown on (100) SrTiO$_{3}$ 
substrates. The low-temperature $\rho$ in the ferromagnetic metallic region follows well 
$\rho (H,T)=\rho _{0}(H)+A(H)\omega_{s}/\sinh (\hbar \omega_{s}/2k_{B}T)+B(H)T^{7/2}$ with 
$\rho _{0}$ being the residual resistivity. We attribute the second and third term to
small-polaron and spin-wave scattering, respectively. Our analysis based on these scattering
mechanisms also gives the observed difference between the metal-insulator transition temperatures
of the films studied. Transport measurements in applied magnetic field further indicate
that spin-wave scattering is a key transport mechanism at low temperatures.
\end{abstract}
\pacs{75.30.-m, 72.10.-d}

\maketitle

The observation of colossal magnetoresistance CMR effect in manganite films \cite{sjin}
has produced a resurgence of interest in these materials for both fundamental physics
and their possible application in recording media and magnetic switching devices. The 
microscopic transport mechanism in these materials has long been thought
to be double exchange DE \cite{zene,ande,kubo}. However, it
has been realized \cite{mill} that the effective carrier-spin interaction in the DE model
is too weak to lead to a significant reduction of the electronic bandwidth, which would
justify the observed several orders of magnitude increase in conductivity just below the
Curie temperature $T_{C}$. Indeed, a large number of experiments have shown that the DE
scenario alone cannot account for the properties of the manganites and that CMR is not
purely electronic in origin \cite{zhao,pdai}.

Low-temperature charge transport measurements of manganites in the ferromagnetic metallic state are
essential in clarifying the specific mechanisms responsible for the CMR effect. At low
temperatures, a dominant $T^{2}$ term in resistivity has generally been observed \cite{urus,snyd,akim}. 
Although the $T^{2}$ behavior is consistent with electron-electron interaction
\cite{babe}, the coefficient of the $T^{2}$ term is about 60 to 70 times larger than the one
expected for electron-electron scattering \cite{kado}. Moreover, a careful check of the low-temperature
resistivity \cite{zhao2,furu} has shown substantial deviation from the $T^{2}-$like behavior in
the very low temperature region. Other power-law temperature dependences of the resistivity have also 
been reported \cite{akim,furu,schi,jaim,zhao3,zhan,cald}. At present, there
is no agreement on the actual scattering mechanism below the Curie temperature.

Here, we address the low-temperature scattering mechanism in manganites through
resistivity measurements of La$_{0.8-y}$Sr$_{0.2}$MnO$_{3}$ ($y=0$ and 0.128) films grown
on SrTiO$_{3}$ substrates, measured in zero field as well as applied magnetic fields up to
14 T. Our data indicate that spin-wave scattering, which gives a $T^{7/2}$ dependence 
in the low-temperature resistivity, is a dominant dissipation mechanism in the ferromagnetic
state of these manganites, besides scattering of small polarons by a soft optical phonon mode. 
Our analysis of the resistivity data in terms of small-polaron and spin-wave scattering
mechanisms, and the spin fluctuation model also gives the observed difference in the 
metal-insulator transition temperature $T_{MI}$ of the three films studied.

Thin films of La$_{0.8}$Sr$_{0.2}$MnO$_{3}$ and La$_{0.672}$Sr$_{0.2}$MnO$_{3}$ were grown
on (100) SrTiO$_{3}$ single crystal substrates by using pulsed laser deposition technique. 
The substrate temperature was 750$^{0}$C, and the oxygen partial pressure in the chamber
was maintained at 0.2 mbar. The growth procedure and its optimization is described elsewhere 
\cite{lebe,chen}. The film thickness was deduced from deposition time normalized to calibration
runs. The DC resistivity of the films was measured in zero-field as well as in applied magnetic 
fields up to 14 T using the standard four-probe technique. The electrical current was in the
film plane, perpendicular to the applied magnetic field. A constant current of 100 $\mu$A provided 
by a Keithley 2400 sourcemeter was used. The magnetization $M$ was measured in a magnetic field
parallel to the film plane using a Quantum Design Superconducting Quantum Interference Device 
(SQUID) magnetometer.

\begin{figure}[tbp]
\begin{center}
\includegraphics[width=\columnwidth]{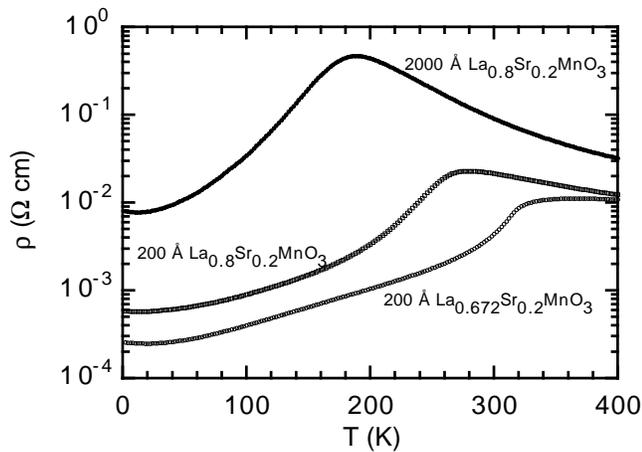}
\end{center}
\caption{Zero-field resistivity $\rho$ as a function of temperature $T$ of the 200 $\AA$
La$_{0.672}$Sr$_{0.2}$MnO$_{3}$ film, and the 200 and 2000 $\AA$ La$_{0.8}$Sr$_{0.2}$MnO$_{3}$ films,
grown on (100) SrTiO$_{3}$ substrates. }
\end{figure}

Figure 1 shows zero-field resistivity data for the 200 $\AA$ lanthanum deficient
La$_{0.672}$Sr$_{0.2}$MnO$_{3}$ film, and for the 200 and 2000 $\AA$ La$_{0.8}$Sr$_{0.2}$MnO$_{3}$ 
films. The 200 $\AA$ lanthanum deficient film has a $T_{MI}$ of 364 K, which is close
to that of a La$_{0.7}$Sr$_{0.3}$MnO$_{3}$ single crystal \cite{urus}. The observed $T_{MI}$ 
of 280 K for the 200 $\AA$ La$_{0.8}$Sr$_{0.2}$MnO$_{3}$ is close to the values
of 270 and 290 K, obtained for as-grown and annealed in N$_{2}$ gas films, respectively, 
with same composition \cite{hlju}. The $T_{MI}$ of 188 K for the 2000 $\AA$
La$_{0.8}$Sr$_{0.2}$MnO$_{3}$ film is lower than that for single crystal specimens \cite{urus}. 
One possible explanation is the nonstoichiometric oxygen content in this film. In fact,
a significant effect of the oxygen content on $T_{MI}$ has been observed in 
La$_{1-x}$Sr$_{x}$MnO$_{3}$ single crystals \cite{shio}. Thus, the composition of this
low-$T_{MI}$ film is probably La$_{0.8}$Sr$_{0.2}$MnO$_{3-\delta}$. The $T_{C}$  of each film, 
determined from magnetization measurements (data not shown), coincides with
its $T_{MI}$. The residual resistivity 
of these films increases with decreasing $T_{MI}$, indicating that the films with lower
$T_{MI}$ are worse metals at low temperatures. 

\begin{figure}[tbp]
\begin{center}
\includegraphics[width=\columnwidth]{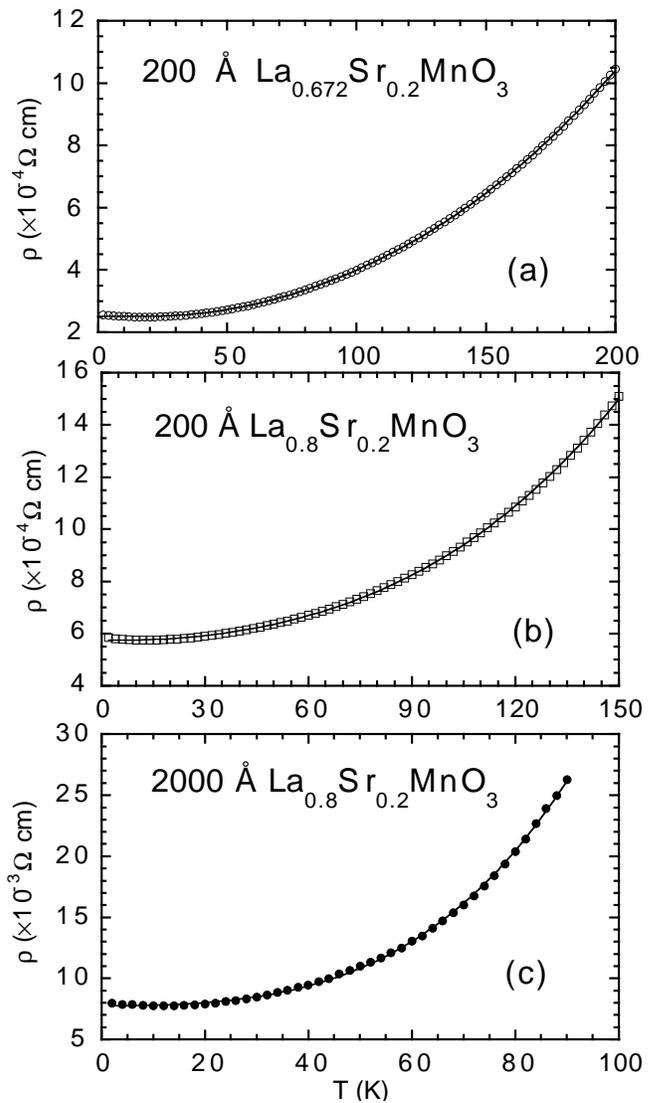}
\end{center}
\caption{ Low-temperature resistivity $\rho (T)$, measured in zero field, for (a) 200 $\AA$
La$_{0.672}$Sr$_{0.2}$MnO$_{3}$ film, (b) 200 $\AA$ La$_{0.8}$Sr$_{0.2}$MnO$_{3}$ film, and
(c) 2000 $\AA$ La$_{0.8}$Sr$_{0.2}$MnO$_{3}$ film, grown on (100) SrTiO$_{3}$ substrates. The
solid lines are fits of the data with Eq. (1). }
\end{figure}

\begin{table*}
\caption{\label{tab:table} Values of the zero field metal-insulator transition temperature $T_{MI}$,
residual resistivity $\rho_0$, fitting coefficients $A$, 
$\hbar \omega _{s}/2k_{B}$, and B, carrier concentration $n$, and activation energy $E_{A}$
of La$_{0.8-y}$Sr$_{0.2}$MnO$_{3}$ films grown on (100) SrTiO$_{3}$ substrates. The definitions 
of the fitting coefficients are given in the text. }
\begin{ruledtabular}
\begin{tabular}{cccccccc}
Samples & $T_{MI}$ (K) & $\rho_{0}$ ($\Omega $ cm) & A ($\Omega $ cm/Hz) & $\hbar \omega 
_{s}/2k_{B}$ (K) & B ($\Omega $ cm/K$^{7/2}$) & $n$ & $E_{A}$ (K) \\
\hline
200 $\AA$ La$_{0.672}$Sr$_{0.2}$MnO$_{3}$ & 364.3 & 2.505$\times 10^{-4}$  & 4.316$\times
10^{-18}$ & 71.93 & 1.669$\times 10^{-12}$ & $>0.2$ & $\sim 1068$\\
200 $\AA$ La$_{0.8}$Sr$_{0.2}$MnO$_{3}$ & 280.2 & 5.747$\times 10^{-4}$ & 2.347$\times
10^{-18}$ & 26.90 & 1.005$\times 10^{-11}$  & $\approx 0.2$ & 1068 \\
2000 $\AA$ La$_{0.8}$Sr$_{0.2}$MnO$_{3-\delta}$ & 188.1 & 7.777$\times 10^{-3}$ & 2.903$\times
10^{-17}$ & 15.11 & 2.077$\times 10^{-9}$  & $<0.2$ & 1546 \\
\end{tabular}
\end{ruledtabular}
\end{table*}

To elucidate the scattering mechanisms in the ferromagnetic metallic region, we plot in 
Fig. 2 the low-temperature behavior of the zero-field resistivity for the
films studied. The resistivities of the three films can be fitted well with
\begin{equation}
\rho (T)=\rho_{0}+\frac{A\omega_{s}}{\sinh ^{2}(\hbar \omega_{s}/2k_{B}T)}+BT^{7/2}~~,
\end{equation}
where $\rho_{0}$, the residual resistivity due to various temperature-independent scattering 
mechanisms, is taken as the resistivity at 10 K, and $A$, $\omega_{s}$ (average frequency of
the softest optical mode), and $B$ 
are fitting coefficients. The excellent fit of the data with Eq. (1) (solid curves in Fig. 2)
suggests that the terms $A\omega_{s}/\sinh ^{2}(\hbar \omega_{s}/2k_{B}T)$ 
and $BT^{7/2}$ capture the basic physics responsible for charge carrier scattering in this
low-temperature region. The values of $T_{MI}$, $\rho_0$, and the fitting coefficients of
these films are given in Table I.

Previous reports have shown that the resistivity of La$_{0.75}$Sr$_{0.25}$MnO$_{3}$ films grown on
(100) LaAlO${_3}$ substrate follows well 
a $\rho(T)$ dependence similar to Eq. (1) in which the third term, however, has a $T^{9/2}$ power-law
dependence, indicative of two-magnon scattering \cite{zhao3}. Nevertheless, the fit of our 
resistivity data with Eq. (1) for the three films studied is better over a wider temperature
range than a fit in which the third term in $\rho(T)$ is $BT^{9/2}$.

According to the theory of small-polaron conduction at low temperatures \cite{lang}, the 
relaxation rate $1/\tau$ for small polarons is proportional to $1/\sinh ^{2}(\hbar
\omega_{s}/2k_{B}T)$ \cite{zhao2,zhao3}. Thus, the term $A\omega_{s}/\sinh ^{2}(\hbar 
\omega_{s}/2k_{B}T)$ in Eq. (1) is consistent with small polaron coherent motion involving
relaxation due to a soft optical phonon mode \cite{zhao2}.  The values of the fitting parameter 
$\hbar \omega_{s}/2k_{B}$ of the La$_{0.8-y}$Sr$_{0.2}$MnO$_{3}$ films are in the 15.1 to
71.9 K range, in agreement with values determined from low-temperature specific heat 
($\hbar \omega_{s}/2k_{B}=$ 48 K) and other resistivity studies of La$_{1-x}$Ca$_{x}$MnO$_{3}$
\cite{zhao2,zhao3}. 
In addition, inelastic neutron scattering \cite{reic} and reflectivity \cite{coul} measurements
support the phononic character of charge carriers for La$_{0.8}$Sr$_{0.2}$MnO$_{3}$. 

We attribute the $T^{7/2}$ dependence of the low-temperature resistivity in Eq. (1) to 
spin-wave scattering. The reason is the following. In the general theory of spin-wave
interactions proposed by Dyson in early 1956, which gives a complete 
description of the thermodynamic properties of a ferromagnet at low temperatures,
the mean-free path $l$ for spin-spin collisions is 
proportional to $T^{-7/2}$ \cite{dyso}. This gives a $T^{7/2}$ temperature dependence for the
resistivity since $\rho=m^{*}v_{F}/(ne^{2}l)$, where $m^{*}$ is the effective mass of the 
charge carriers, $n$ is the carrier concentration, and $v_{F}$ is the Fermi velocity. In fact,
when the spin orientation angle between neighboring sites is small enough, the DE Hamiltonian 
in the low-temperature ferromagnetic state can be mapped into the Heisenberg Hamiltonian.

\begin{figure}[b]
\begin{center}
\includegraphics[width=\columnwidth]{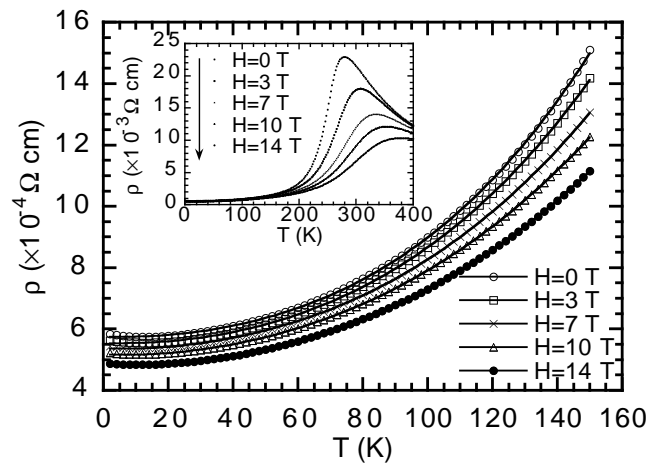}
\end{center}
\caption{Low-temperature resistivity $\rho (T)$ of the 200 $\AA$ La$_{0.8}$Sr$_{0.2}$MnO$_{3}$
film grown on a (100) SrTiO$_{3}$ substrate, measured in various applied magnetic fields $H$. The solid 
lines are the fits of the data with Eq. (1). The inset shows $\rho$ vs $T$ for the same film
over the whole measured $T$ range at various $H$. }
\end{figure}

The temperature independent term $\rho_{0}$ in Eq. (1) is usually ascribed to scattering
from impurities, defects, grain boundaries, and domain walls. The values of 
$\rho_{0}$ for the 200 $\AA$ La$_{0.8-y}$Sr$_{0.2}$MnO$_{3}$ films are comparable with the
value for a La$_{0.8}$Sr$_{0.2}$MnO$_{3}$ single crystal \cite{urus}, indicating 
weak external scattering. In general, $\rho_{0}$ is proportional to $m^{*}/n\tau_{0}$, where
$\tau_{0}$ is the zero-temperature relaxation time. Assuming that $\tau_{0}$ does not change 
a lot among the present films, the smaller (larger) value of $\rho_{0}$ for the 200 $\AA$
La$_{0.672}$Sr$_{0.2}$MnO$_{3}$ film (2000 $\AA$ La$_{0.8}$Sr$_{0.2}$MnO$_{3}$ film) 
than for the 200 $\AA$ La$_{0.8}$Sr$_{0.2}$MnO$_{3}$ film indicates that $n> 0.2$ ($n< 0.2$).
As discussed above, $n< 0.2$ is probably a result of oxygen vacancies. These results for $n$ 
are consistent with the defect chemistry \cite{roos} and are included in Table I.

Next we show that the observed difference in $T_{MI}$ or $T_{C}$ for the three films studied
can be explained based on the values of the above fitting parameters. The spin 
fluctuation model \cite{varm} gives $T_{C}\simeq Wn(1-n)/20$, where $W$ is the electronic
``bare'' bandwidth. However, it has been shown that the huge isotope effect \cite{zhao}, 
the strong sensitivity to oxygen content \cite{dete}, and the significant strain effect \cite{chen}
present in these manganites can be well explained if $W$ is replaced by an effective bandwidth 
$W_{eff} \propto W
\exp(-\gamma E_{b}/\hbar \omega)$, where $E_{b}$ is the binding energy of the polarons, 
which can be estimated from the activation energy $E_{A}$ as $E_{b}\simeq 2E_{A}$,
$\omega$ is the characteristic frequency of the optical phonon mode, which can be taken 
as $\omega\simeq \omega_{s}$, and $\gamma$ is a positive constant. The expression of $T_{C}$,
in which $W$ is replaced by 
$W_{eff}$, shows that $T_{C}$ increases as $W$, $\omega$,
and $n$ increase (for $n\leq 0.5$) and $E_{A}$ decreases. 
Spin-wave scattering implies that the fitting coefficient $B\propto D_{s}^{-7/2}$ \cite{furu},
where $D_{s}\propto W$ is the spin-wave stiffness coefficient \cite{ande,kubo}. 
Hence, an increase in $W$ is reflected as a decrease in $B$. We determined $E_{A}$ to be 1068 and
1546 K for the 200 and 2000 $\AA$ La$_{0.8}$Sr$_{0.2}$MnO$_{3}$ film, respectively, by 
fitting the resistivity data in the high-temperature paramagnetic region in terms of the
adiabatic small-polaron model \cite{emin}. We took $E_{A}$ for the 200 $\AA$ 
La$_{0.672}$Sr$_{0.2}$MnO$_{3}$ film, for which the available data in the paramagnetic region
are over a narrow temperature range due to its higher $T_{MI}$, to be the same as $E_{A}$ for 
the 200 $\AA$ La$_{0.8}$Sr$_{0.2}$MnO$_{3}$ film \cite{chen}. These values of $E_{A}$ are
also included in Table I. As predicted above, Table I shows that, indeed, the film with larger 
$\omega$ and $n$ and smaller $B$ and $E_{A}$ has higher $T_{MI}$. Therefore, the present results
further provide strong support for the small phonon and magnon scatterings as 
low-temperature dissipation mechanisms.

\begin{figure}[tbp]
\begin{center}
\includegraphics[width=\columnwidth]{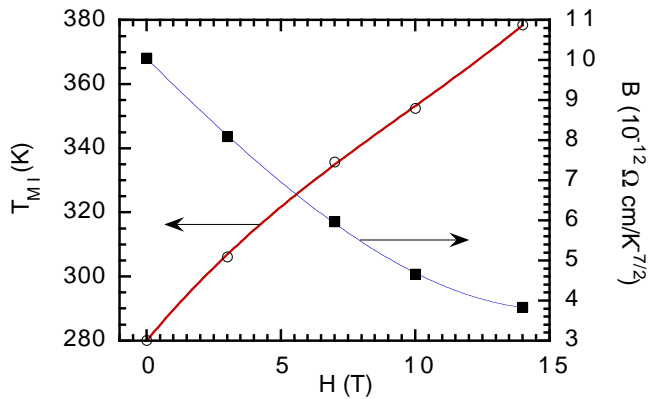}
\end{center}
\caption{ Magnetic field $H$ dependence of both the fitting parameter $B$ (closed squares) of Eq. (1)
and the metal-insulator transition temperature
$T_{MI}$ (closed circles) for the 
200 $\AA $ La$_{0.8}$Sr$_{0.2}$MnO$_{3}$ film grown on a (100) SrTiO$_{3}$ substrate. The lines are
guides to the eye. }
\end{figure}

We now address the effect of the magnetic field on the low-temperature conduction.
Figure 3 is a plot of $\rho$ vs $T$ data ($T\leq 120$ K) of the 200 $\AA$
La$_{0.8}$Sr$_{0.2}$MnO$_{3}$ film measured at various magnetic fields, while its
inset shows $\rho (H,T)$ over the whole measured temperature range ($2\leq T \leq 400$ K). 
Fits of the main panel data with Eq. (1), with A and B fitting parameters and $\omega_s$
taken from zero-field fitting, give the solid curves. 
Notice the excellent agreement between the curves and the data. Equally good fitting
results were obtained for the other two films studied. This universality of charge 
dissipation at low temperatures with respect to thickness, composition, and magnetic
field further indicates that the proposed dissipation mechanisms are intrinsic.

The above fitting has shown that $A$ is only weak field dependent, while $B$
has a stronger field dependence [$B(H)$ is shown in Fig. 4]. Hence, the magnetoresistance
observed in this low-temperature range is absorbed primarily in the $T^{7/2}$ term. This 
is a reasonable result since this term is the spin-wave contribution to
scattering, hence, resistivity.

Figure 4 shows the magnetic field dependence of both $B$ and $T_{MI}$ of the 200 $\AA$ 
La$_{0.8}$Sr$_{0.2}$MnO$_{3}$ film. The effect of an applied field is to open an energy
gap in the magnon spectrum. As a result, the spin-wave scattering should decrease with 
increasing $H$. This is, indeed, reflected by the decrease of $B$ with increasing $H$.
Note also that $T_{MI}$ increases with increasing $H$. This correlation between $T_{MI}$ 
and $B$ further indicates that spin-wave scattering is a key transport mechanism at
low temperatures in these manganites.

In conclusion, we presented resistivity measurements in fields up to 14 T on 
La$_{0.8-y}$Sr$_{0.2}$MnO$_{3}$ films with different composition and/or thickness. Our results
indicate that the dissipation mechanisms in the low-temperature ferromagnetic state are spin-wave 
and small-polaron scattering. The fitting parameters obtained by analyzing the resistivity data
in terms of these dissipation mechanisms explain the observed difference in $T_{MI}$ among the films 
studied. Transport measurements in applied magnetic fields indicate that spin-wave scattering
is an essential dissipative mechanism at low temperatures.

This research was supported at KSU by the National Science Foundation under Grant No.
DMR-0102415.

\end{document}